\documentclass{article}
\usepackage[margin=1in]{geometry} 

\usepackage{graphicx}

\usepackage{amsmath}
\usepackage{amsthm}
\usepackage{amssymb}

\usepackage{subcaption}
\title{Real-time reconstruction of intense, ultrafast laser pulses using deep learning}

\author{Matthew Stanfield$^{*,1}$ \and Jordan Ott$^{*,2}$ \and Christopher Gardner$^{1}$ \and Nicholas F. Beier$^{1}$ \and Deano Farinella$^{1}$ \and Christopher A. Mancuso$^{3}$ \and Pierre Baldi$^{\dagger,2}$ \and Franklin Dollar$^{\dagger,1}$}
\date{%
1. STROBE, NSF Science \& Technology Center, University of California, Irvine, CA 92617\\
2. Department of Computer Science, University of California, Irvine, CA 92617\\
3. Department of Computational Mathematics Science and Engineering, Michigan State University, East Lansing, MI 48824 \\
* These authors contributed equally.\\
$\dagger$ Corresponding authors: \{stanfiem,pfbaldi,fdollar\}@uci.edu
}

\begin{document}
\maketitle

\begin{abstract}
Ultrafast lasers ($< 500$ fs) have enabled laser-matter interactions at intensities exceeding $10^{18} \rm{Wcm}^{-2}$ with only millijoules of laser energy. However, as pulse durations become shorter, larger spectral bandwidths are required. Increasing the bandwidth causes the temporal structure to be increasingly sensitive to spectral phase, yet measuring the spectral phase of a laser pulse is nontrivial. While direct measurements of the spectral phase cannot be done using square-integrable detectors, phase information can be reconstructed by measuring the spectral response of a nonlinear optical effect. We introduce a new deep learning approach using the generalized nonlinear Schr\"{o}dinger equation and self-phase modulation, a $\chi_3$ nonlinearity occurring from material propagation. By training a neural network on numerical simulations of pulses propagating in a known material, the features of spectral change can be use to reconstruct the spectral phase. The technique is also sensitive to the local fluence of the pulse, enabling the full temporal intensity profile to be calculated. We demonstrate our method on a simulated large bandwidth pulse undergoing moderate material dispersion, and an experimentally produced broadband spectrum with substantial material dispersion. Error rates are low, even when modest amounts of noise introduced. With a single plate of glass and an optical spectrometer, single shot phase and fluence measurements are possible in real-time on intense ultrafast laser systems.  
\end{abstract}

\section{Introduction}

High intensity ultrafast laser systems ($< 500$ fs) are used for a wide variety of applications, such as x-ray generation \cite{Naumova2004,Heissler2012}, electron acceleration \cite{Tajima1979,Faure2004,Mangles2004,Geddes2004}, and ion acceleration  \cite{torrisi2019protons,zhou_proton_2016}. The short pulse duration is the key aspect in delivering high peak intensity bursts, even if the overall energy of the pulse is relatively low. Creating a shorter pulse duration requires increasing the spectral width of the pulse along with having minimal phase shifts between the frequencies. If significant phase differences exist in the pulse, the pulse duration can drastically increase, significantly dropping the peak intensity of the pulse. While knowing the temporal profile is essential for many applications, measuring the temporal profile is non-trivial. The duration of a single optical cycle is on the order of a few femtoseconds, orders of magnitude quicker than the rise times of the fastest electronics. 

Numerous techniques exist to measure the temporal profile of an ultrafast laser pulse \cite{trebino2020highly, Iaconis:98, Fittinghoff:96, trebino1997measuring, miranda2012simultaneous, prade_simple_1994, nibbering_spectral_1996, anashkina_single-shot_2016,OShea:01}. A nonlinear interaction is typically used to encode the phase information into a signal measurable by a square-integrable detector. Techniques such as intensity auto-correlation and frequency resolved optical gating (FROG) rely on measuring a nonlinearity induced from the interaction between two or more pulses \cite{trebino1997measuring}. Other techniques, such as dispersion scan (D-Scan), rely on changing the phase of the initial pulse by a known amount and monitoring how that affects the nonlinear interaction \cite{miranda2012simultaneous}. Commonly, a second harmonic mechanism is used for the nonlinear effect but other nonlinearities, such as the effects originating from the third-order term of the nonlinear electric susceptibility, $\chi_3$, have been used \cite{trebino2020highly,Tsang:96}. Often FROG and D-Scan are used in scanning geometries, requiring many shots from a stable laser. 

Another nonlinearity that can be used for pulse measurement is self-phase modulation (SPM). SPM is a nonlinear optical effect that occurs due to an intensity dependent index of refraction called the optical Kerr effect, which is a $\chi_3$ effect \cite{shimizu1967frequency}. The nonlinear change in index takes the form of $n = n_0 + \gamma P(\tau)$, where $n_0$ is the linear index of refraction, $\gamma$ is the nonlinear coefficient, and $P(\tau)$ is the temporal power profile of the pulse. SPM can be modeled by the generalized nonlinear Schr\"{o}dinger equation (GNLSE), which takes into account the effects of material dispersion, delayed Raman effect, and  self-steepening \cite{hult}. If these effects are negligible, then the GNLSE is able to be solved analytically, taking the form of a nonlinear temporal phase shift, $E(z) = E(0)e^{i\gamma P(\tau)z}$. The total amount of nonlinearity of the system is described by the B-integral, which is the integral of the nonlinear phase shift accumulated through self-phase modulation. If the temporal profile does not change during propagation, the B-integral simplifies to $B = \gamma P_{max}z$ with $P_{max}$ being the peak power of the pulse. If material dispersion, delayed Raman effect, and self-steepening are non-negligible, SPM is no longer well modeled by the analytical solution to the GNLSE and requires numerical methods to accurately model the spectral changes. 
\begin{figure}[t]
\centering
\includegraphics[width=\linewidth]{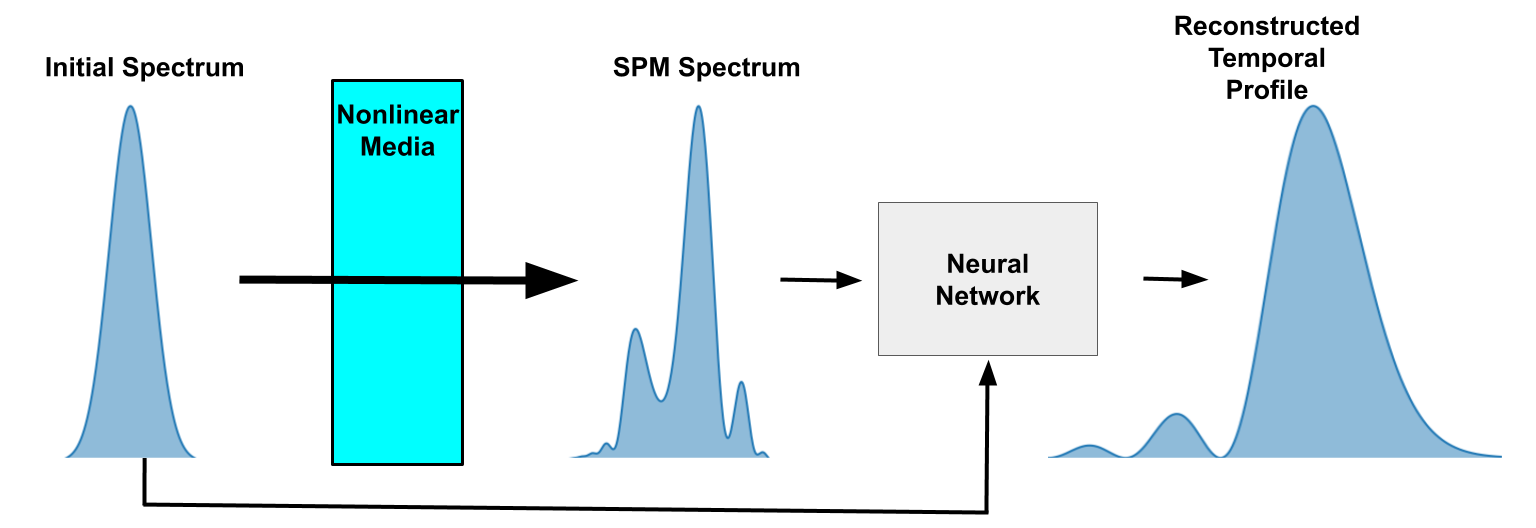}
\caption{Example setup for using self-phase modulation to measure the initial temporal profile of a pulse. The initial spectrum passes through a material with a Kerr nonlinearity, causing a change in the spectrum due to self-phase modulation. A neural network takes the initial and final spectrum as inputs and extracts the initial phase and fluence of the initial pulse, allowing the initial temporal profile to be reconstructed.}
\label{fig:setup}
\end{figure}
Since the spectral change due to SPM is directly dependent on the temporal power, both the spectral phase and peak power can be reconstructed from the change in the laser spectrum without any direction of time ambiguity (i.e. which side of the pulse is the leading edge of the pulse). Combining the peak power reconstruction with knowledge of the spatial profile of the laser pulse, the peak intensity of the laser pulse can also be calculated. Commonly, the analytical solution for the GNLSE has been used in an Gerchberg-Saxton style iterative phase reconstruction algorithm, in which the phase is reconstructed from the measured spectra of subsequent thin dielectric plates  \cite{prade_simple_1994,nibbering_spectral_1996,anashkina_single-shot_2016}.

Iterative methods have seen success for many-cycle laser pulses, however relying on the analytical solutions to the GNLSE limits the application to systems with negligible material dispersion. For broadband laser systems, even a small amount of material dispersion can substantially alter the way the spectrum changes from SPM, requiring the GNLSE to be solved numerically.

Deep learning based algorithms recently have shown great promise for ultrafast laser pulse reconstructions. Deep learning has been applied to other pulse measurement techniques, such as SHG FROG and D-Scan, where neural networks replaced iterative algorithms \cite{zahavy_deep_2018,kleinert2019rapid}. Deep learning can directly learn nonlinear relationships between various features within data and map them to the desired target variables \cite{baldi2021deep}. Since the information is present in the data used in the iterative phase reconstruction algorithms, deep learning can be used to directly learn the transformation between the data and the reconstructions without the need for the iterative algorithms. Since deep learning methods bypass the need for an iterative algorithm during reconstruction, deep learning approaches can be significantly faster than their iterative counterparts, enabling real-time reconstructions.

By utilizing the feature learning of neural networks, the full nonlinear propagation of the light through the material can be learned. This enables phase reconstruction to occur even if material dispersion is non-negligible, which is nearly unavoidable for broadband few-cycle laser pulses. The neural networks only requires information about the spectrum of the laser before and after SPM, as shown in Fig. \ref{fig:setup}. For laser systems with a stable initial spectral profile, this information can be assumed while training, removing the requirement of it being included in the features for the neural network. Otherwise the initial spectrum is required to be included as a feature that is passed to the neural network.  

In this study, we demonstrate a robust deep learning reconstruction of the spectral phase and peak fluence of ultrafast laser pulses from the spectral changes imparted by self-phase modulation. The reconstructions coupled with the spatial profile of the laser system enable the peak intensity to be inferred. Neural networks were trained separately for two different systems, a simulated broadband laser source and an experimental Ti:Sapp laser source. For the simulated broadband laser source two neural networks were trained to predict the initial phase and the fluence of the laser pulse. Due to the stability of initial laser fluence and spectral profile of the experimental system, only a single network trained on the SPM spectrum was needed to predict the initial phase of the laser. All neural networks were trained on simulated data generated from numerically solving the GNLSE to model the nonlinear propagation of a laser pulse through a dielectric media. Applications to experimental measurements is demonstrated with phase measurements of a 800 nm Ti:Sapphire laser being retrievable in real-time.

\section{Broadband Simulation Results}
\begin{figure}[t]
\centering
\includegraphics[width=\linewidth]{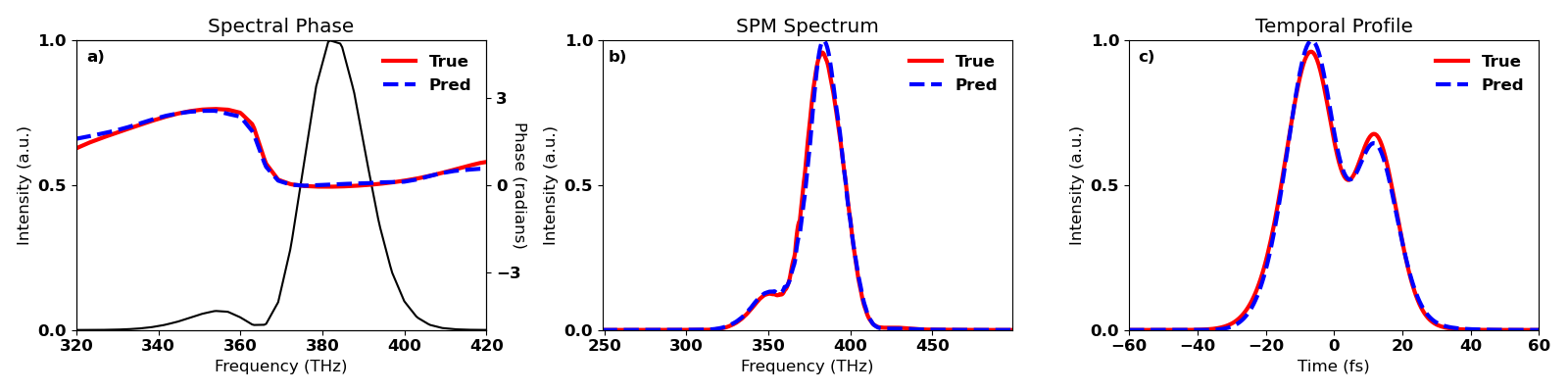}
\caption{A) An example of a randomly generated spectrum (black) centered on 374 THz along with the randomly generated phase (red).B) GNLSE simulations based from the reconstructed initial pulse(dashed blue) and the true initial pulse (red). C) Temporal profiles of the reconstructed pulse(dashed blue) and the true temporal profile (red). }
\label{fig:single}
\end{figure}

\begin{figure}[htbp]
\centering
\includegraphics[width=\linewidth]{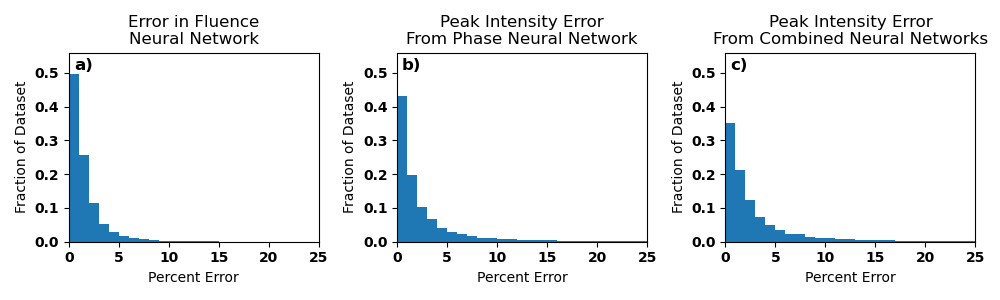}
\caption{\textbf{Reconstruction Error in Peak Intensity Prediction} To show the accuracy of the reconstructions of the neural networks the percent error is shown. a) Reconstruction error in the predictions for the fluence of the pulse. b) Reconstruction error in the predictions for the maximum of the normalized temporal profile of the reconstructed pulse. c) Reconstruction error in the predictions for the peak intensity of the reconstructed pulse using predicted fluence of the pulse. }
\label{fig:error}
\end{figure}
For the broadband simulated spectra, two separate neural networks were trained on randomly generated simulated pulses. Both networks were designed to make predictions off of the spectral measurements of the initial pulse and the pulse after SPM, with one network used to predict the initial phase of the pulses and the other network used to predict the initial fluence of the pulses. After training, 20,000 pulses withheld from the training data were run through the networks to test the accuracy of the reconstructions for previously unseen data. To quantify the accuracy of the neural network on the physical qualities we are predicting, the relative reconstruction error is calculated for the fluence and the peak value of the normalized temporal profile. To quantify the combined accuracy of the two networks the peak value of the intensity profile is calculated.

\begin{figure}[htbp]
\centering
\includegraphics[width=\linewidth]{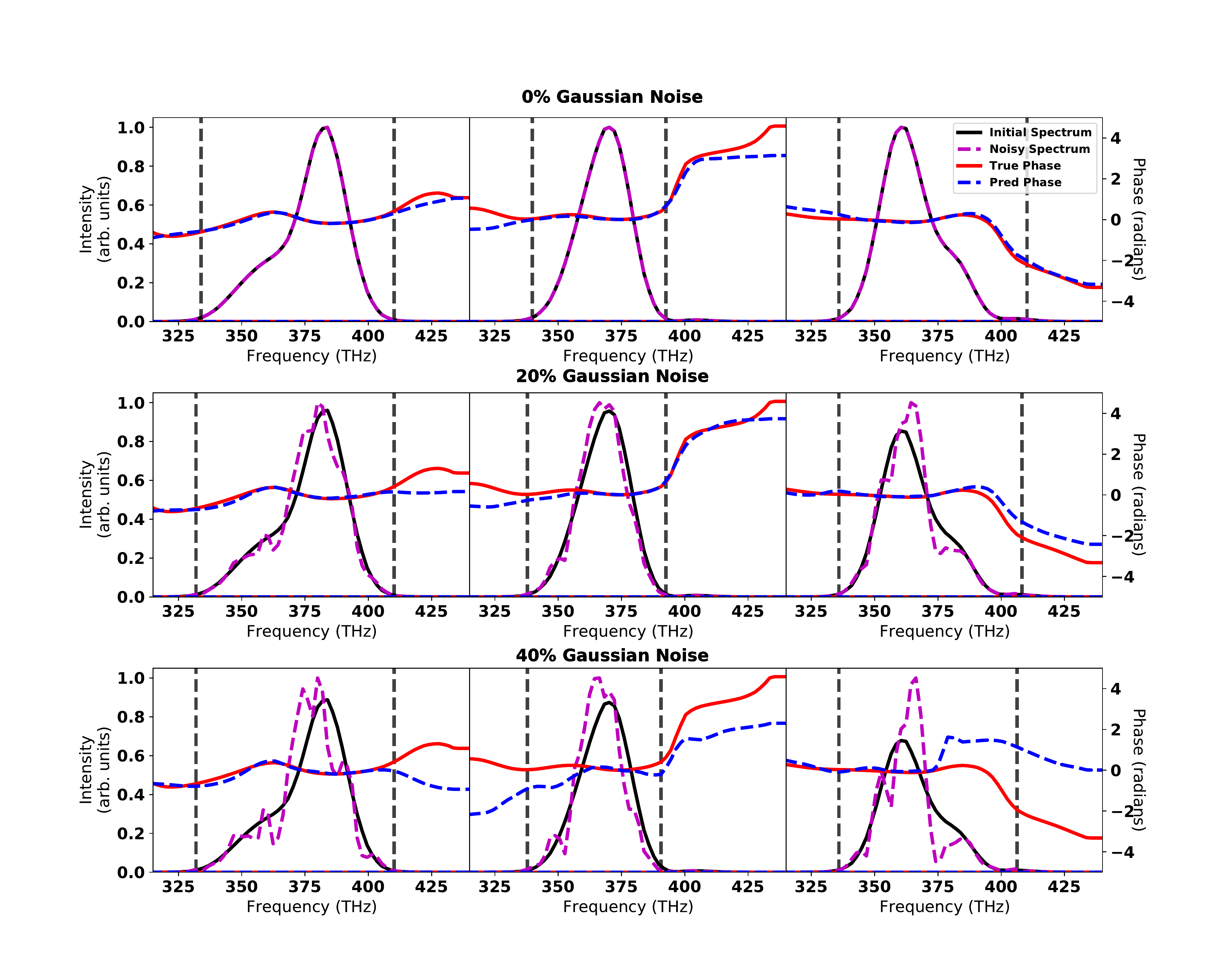}
\caption{\textbf{Reconstructions with Gaussian Noise} To show the robustness of the phase predictions from the neural network  above is three random initial spectra(black) and the noisy spectra(magenta) after adding a 0\%, 20\%, and 40\% Gaussian noise. The true phase (solid red) is shown in comparison  to the noisy phase (blue dashed) predicted from the noisy spectrum. The vertical dashed lines denote the location where the spectral intensity falls below 1\% of the maximum value, outside this region the phase is ill-defined.}
\label{fig:noisy_spectrum_scaling}
\end{figure}

Comparing the reconstructed fluence to its known value provides a way to measure the accuracy of the neural network's predictions on the physical values we are trying to predict. For $99\%$ of the pulses in the test data, the neural network was able to predict the peak fluence within an error of $<10\%$, a mean fluence reconstruction error of $1.6\%$ and a standard deviation of the fluence reconstruction error being $2.1\%$[Fig \ref{fig:error}]. When calculating the error of the phase reconstruction, we only considered regions within the pulses dB-20 spectral width, since the phase is ill-defined and not physically meaningful outside of areas with significant power spectrum. The mean standard deviation of  the predicted phase was 0.13 radians. To examine how the accuracy of the phase neural network translates into the temporal domain, the reconstruction error in the predicted maximum of the normalized temporal profile was calculated. For this calculation,  $93\%$ of the pulses had a reconstruction error below $10\%$, a mean reconstruction error of $3.3\%$ and a standard deviation of the error of $6.7\%$. An example of a reconstructed pulse is shown in Fig. \ref{fig:single}.

By combining the results from the fluence neural network with the results of the phase neural network, the temporal intensity profile can be reconstructed, including the direction of time. Using the peak temporal intensity as an estimate of the intensity reconstruction error, which has an mean intensity reconstruction error of $3.7\%$ and a standard deviation of the intensity reconstruction error of $7.1\%$  with over $90\%$ of the data set has less than $10\%$ error.   

Even in the presence of noise, accurate reconstructions can be obtained. In Fig. \ref{fig:noisy_spectrum_scaling}, three pulses are shown with their phase reconstructions in the presence of 0\%, 20\%, and 40\% Gaussian noise. When applied to the entire test data set, the 20\% Gaussian noise caused an increase of the mean intensity reconstruction error to $5.9\%$ and a standard deviation of the intensity reconstruction error of $8.3\%$  with over $85\%$ of the data set has less than $10\%$ error. For 40\% Gaussian noise the mean intensity reconstruction error to $50.1\%$ and a standard deviation of the intensity reconstruction error of $61.2\%$  with $25\%$ of the data set has less than $10\%$ error.

\section{Experimental Results}

\begin{figure}[htbp]
\centering
\includegraphics[width=0.75\linewidth]{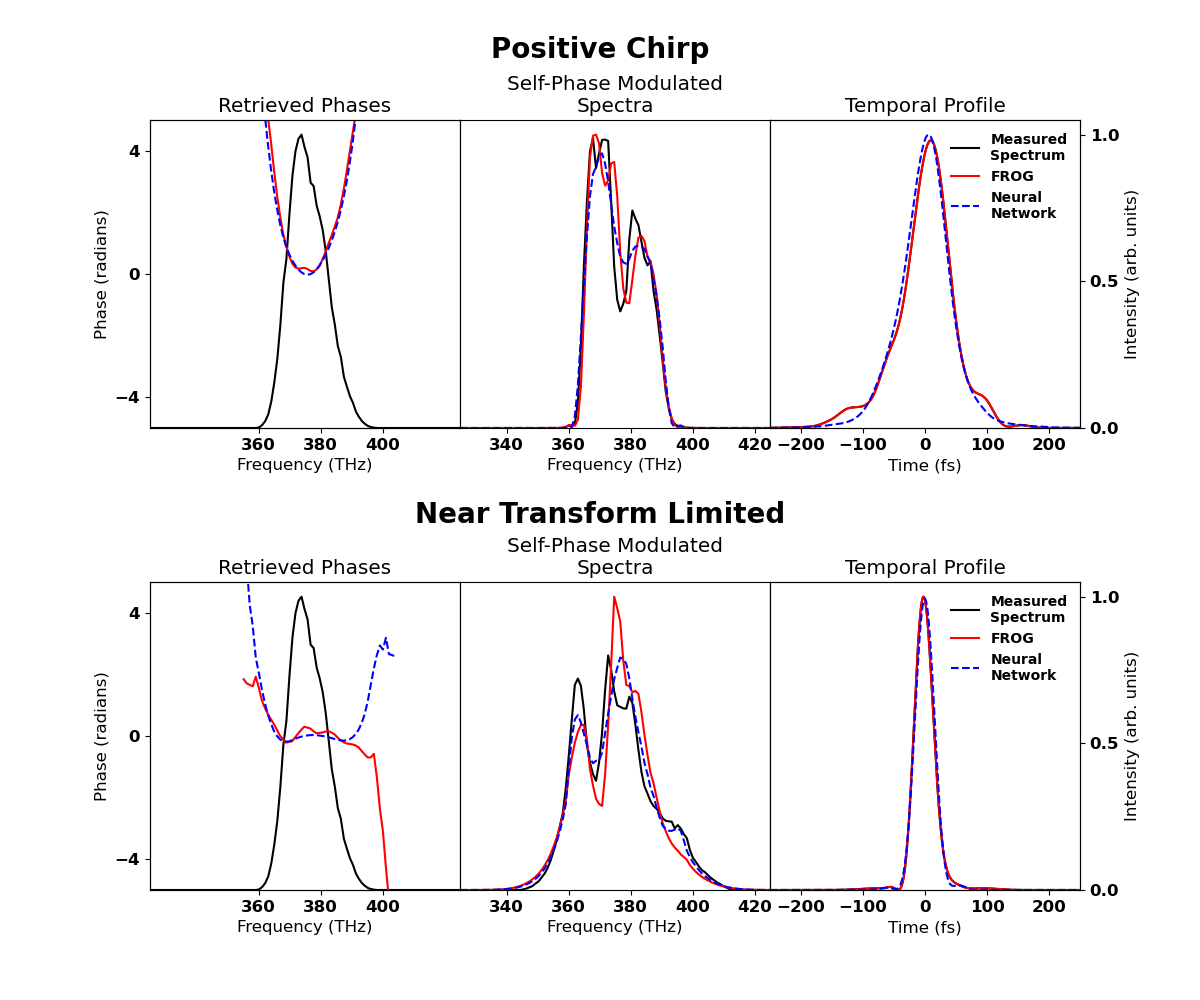}
\caption{\textbf{Experimental Phase Reconstruction} Phases measured from experimental data. Spectra measured after propagating through 8 mm of fused silica at Brewster's angle. Spectra and temporal profiles are normalized to the area under the curve. a) FROG reconstruction of a positively chirped laser pulse in comparison to the reconstructed results from the neural network. Self-Phase modulated spectra and temporal profile are normalized to area under the curve for ease of comparison. b) FROG reconstruction of a near transform-limited pulse in comparison to the reconstructed results from the neural network. }
\label{fig:experimental}
\end{figure}

In additional to validating the technique on broadband simulated data, we also trained a separate neural network to make predictions on experimental data. This experiment was performed on a commercially available, 1 kHz repetition-rate laser system (Spectra-Physics Solstice ACE) with an energy of 6.6 mJ, beam diameter of 12 mm, central wavelength of 800 nm, and FTL pulse duration of 34 fs. The output of the laser was characterized using an SHG FROG. The nonlinear media, 8 mm of fused silica, is oriented at Brewster's angle such that the effective propagation length after taking into account refraction is 9.6 mm. The collimated laser beam has a peak fluence of 11.7 mJcm$^{-2}$. Due to the large amount of material that the laser is propagating through, material dispersion will significantly impact the SPM spectrum.  The spectra were taken by isolating the center of the beam with a hard aperture and sent to a fiber spectrometer utilizing a cosine corrector to minimize spectral interference from occurring inside the optical fiber. 

Due to the laser having $<0.5\%$ root mean square energy fluctuations and a standard deviation of transform limited pulse duration $<0.75$ fs, the initial laser spectrum can be assumed to be constant and is not needed to be included in the features given to the neural network. The reconstruction of the experimental neural network for two separate pulses is shown in Fig. \ref{fig:experimental}. The pulse duration for the positive chirped pulse predicted by the neural network was 73 fs compared to the FROGs 76 fs. The pulse duration predicted by the neural network for the near transform-limited pulse was 36 fs compared to the FROGs 35 fs pulse duration.

\section{Discussion}

With this work, we have shown an inexpensive and easy to experimentally implement method for measuring the temporal intensity profile of the laser pulse by utilizing self-phase modulation. When combined with the knowledge of the focal spot of an experimental system, this information would enable peak intensity to be calculated. Since the technique does not rely on any scanning for the measurement, this can enable single-shot intensity profile characterization of the laser.

When applied to experimental data, as shown in \ref{fig:experimental}, we see that both the FROG and the neural network method are in good agreement with the SPM spectrum, and the reconstructed temporal profiles are nearly identical. Since the spectral modulations tend to not contain fast varying features, high spectral resolution is not needed. For example, the neural network trained on simulated broadband pulses was trained on data with a wavelength resolution of $>4$ nm per pixel. The phase was able to be reconstructed using the neural network in under 10 ms on a commercially available desktop computer, implying that real-time phase reconstruction is possible.

While two separate neural networks were trained to predict the phase for the initial pulse, this was only done to simplify the data generation process. Since each neural network is only able to predict pulses of similar structure to what it was trained on, a specifically tailored network was designed for the few-cycle simulated pulses and many-cycle experimental pulses. A single network could have been trained, but would require the training data to span a wide range bandwidths and spectral phases while maintaining an intensity large enough to introduce undergo SPM. By training different networks, the total amount of data that needs to be generated can be reduced.

While this work is based on calculating the intensity and pulse duration of a pulse at a single location in the beam, having the temporal information as a function of a beam's spatial profile is possible by sampling the spectral broadening at multiple locations in the beam, building a 3-dimensional intensity mapping of the beam. Large aperture beams, such as the many centimeter diameter beams at multi-petawatt laser systems such as ELI or ZEUS are often measured in a small subset of the beam, typically at low energy. Using an imaging spectrometer that can raster across the transverse beam spatial profile, this method can completely characterize the beam.  

OPCPA laser systems have enabled high power few-cycle laser systems in the mid-infrared (MIR) wavelength regions. In these regions traditional silicon-based detectors no longer work, meaning forcing a reliance on more expensive InGaAs detectors. While this technique was demonstrated for wavelengths from a Ti:Sapphire laser system, the technique could be applied to the wavelengths in the MIR region.  Scaling to other wavelengths would only require knowledge of material properties of the nonlinear media used along with retraining of the neural network. Being able to reconstruct the phase from only two spectral measurements enables the phase information to be readily obtainable from a field auto-correlator, meaning phase information could be reconstructed from a single power diode. Field auto-correlators are already commonly used for techniques such as Fourier transform infrared spectroscopy, making this technique simple to implement into such systems. 

The ability for self-phase modulation to spectrally broaden a pulse is utilized in many pulse compression techniques to generate a few-cycle laser pulses. Modifying the neural network to predict the spectral phase after self-phase modulation would enable the reconstruction of the temporal profile after pulse compression. With this modification, the same system could be used to generate and characterize a few-cycle laser pulse. 

\section{Conclusion}
The presented pulse measurement technique shows a general technique of measuring the intensity profile of a laser pulse in single-shot applications using inexpensive and readily available components, only requiring a piece of glass and a spectrometer. By using a fully connected neural network phase reconstruction based on the generalized nonlinear Schr\"{o}dinger equation is able to be done, which includes material dispersion, delayed Raman effect, and  self-steepening. Since material dispersion is included in the modeling this technique is able to be used to characterize broadband few-cycle laser pulses in real-time. With minor modifications, this technique enables measuring the fluence and spectral phase of the pulse across the wavefront, enabling measuring variance in the temporal profile across the beam for large aperture beams that are common at facilities such as ZEUS.

\section{Methods}
\label{Methods}

\subsection{Numerical Data Generation}
\label{datagen}

In order to generate sufficiently large data sets for training neural networks, numerical simulations of a wide range of ultrafast laser pulses with varying phases were produced. The nonlinear propagation could then be modeled and the nonlinear propagation of those pulses was modeled. The simulations were performed with PyNLO, a python based 1-dimensional GNLSE solver using the split-step Fourier method \cite{hult,2015pynlo}. PyNLO numerically models material dispersion, self-steepening, and the delayed Raman response. The central frequency of the simulations was set to 374.0 THz, which is the central frequency of Ti:Sapphire lasers, the most common ultrafast laser. The material properties were based on the values for fused silica, which is a common optical glass that is able to be obtained with high optical quality and is well characterized. The material dispersion was modeled by using the second, third, and fourth order expansion curves of the Sellmeier equation for fused silica, which are $36.1$ fs$^2$mm$^{-1}$, $27.49$ fs$^3$mm$^{-1}$, and $-11.4335$ fs$^4$mm$^{-1}$. 

Spatial effects , such as beam breakup or self-focusing, may experimentally limit the technique to B-integrals of $<3$. This is due to spatial effects will break the 1-dimensional assumption of the underlying model  \cite{CUMBERBATCH_1970,hult}. An average B-integral of $~2$ was desired for the simulation, as this will ensure enough nonlinearity to induce a strong spectral response while maintaining a B-integral reasonable for experiments. 

\subsubsection*{Broadband Simulated Data Generation}

To ensure a representative set of phases and spectrum were present, the training data was generated from a randomly generated vector. The vector has a Gaussian envelope applied in the temporal and spectral domains, generating a pulse with a random spectrum and spectral phase. The temporal envelope used to generate the data is 30 fs. The spectral envelope used has a width of 40 THz centered on 374 THz.

After the temporal and spectral envelopes have been applied, the peak fluence is set by randomly sampling from a uniform distribution spanning the range of 16.2 mJcm$^{-2}$ to 43.2 mJcm$^{-2}$. To remove the constant phase ambiguity, the spectral phase was defined to be zero at the central frequency. To remove the linear phase ambiguity, the temporal power's central moment was set to be centered at t=0. Simulations were then ran using PyNLO inside of 1 mm of fused silica assuming a nonlinear coefficient, $\gamma$, of $6$x$10^{-8}$ $ (Wm)^{-1}$. Due to the method of generating random spectra, some pulses with B-integrals $>3$ are generated. These pulse are not filtered out but experimental pulses with this B-integral may run into spatial effects that break the 1-dimensional assumptions made in this work. The resulting simulations had a B-integral ranging from $~0.65$ to $~4.25$, with an average B-integral of $2.23$.  

The initial phase and initial spectrum were interpolated to 40 linearly space bins spanning a frequency range of 120 THz centered on 374 THz. The SPM spectrum was interpolated onto a linearly spaced vector with 100 bins and spanning the frequency range of 300 THz centered on 374 THz. After interpolation the area under the curve (i.e. the energy) for both the initial and SPM spectrum were normalized to unity to ensure the neural network is learning from the relative shape changes of the spectra. The initial and SPM spectra are then combined to create the feature vector that the neural network is trained on. A total of 1,830,000 samples were generated for the training and validation sets, with an additional 20,000 samples generated for the test set.

\subsubsection*{Experimental Data Generation}
To train this network, simulations used for data generation is based on the experimentally measured initial laser spectrum of the laser.  Due to the dominate phase terms of the pulse being group delay dispersion (GDD) and third-order dispersion (TOD), the phase was modeled primarily as a Taylor series expansion, with random GDD, TOD, and forth-order (FOD) phase terms. The dispersion coefficient were generated from a normal distribution with the standard deviation of 10$^3$ fs$^2$, 10$^4$ fs$^3$, 10$^6$ fs$^4$ for the GDD, TOD, and FOD phase terms. To allow for minor deviations from this expansion, a random phase was generated by taking the phase of a random spectrum generated using the Fourier technique described used to generate the broadband simulated data and was added to the Taylor series phase.  This Fourier phase was generated using a temporal and frequency FWHMs used were 60 fs and 50 THz with a maximum phase deviation within 25 THz of the central frequency being sampled from a normal distribution with a standard deviation of $0.5\pi$.  The peak fluence of the pulse was set to match the fluence from the laser and propagated through 9.6 mm of fused silica. A total of 432 thousand simulated pulses were used for training the network.

\subsection{Neural Networks}
\label{Neural Network}

\subsubsection*{Broadband Simulated Neural Networks}

\begin{table}[]
    \centering
    \begin{tabular}{c|c|c|c|c|c}
        Name & Range & Parameter Type & Phase & Fluence & Baseline \\ 
        \hline
        Batch Normalization & (yes, no)             & Choice & yes & yes & no \\
        Dropout             & (0., 0.25)            & Continuous  & 0.175 & 0. & 0. \\
        Learning Rate       & (0.00001, 0.01)       & Continuous (log)  & 0.008 & 0.003 & 0.001\\
        Learning Rate Decay & (0.5, 1.)             & Continuous  & 0.98 & 1. & 1. \\
        Number of Layers    & (3, 20)               & Discrete  & 4 & 8 & 5\\
        Number of Nodes     & (128, 512)            & Discrete  & 505 & 360 & 256\\
        Optimizer           & (Adam, SGD, RMSProp)  & Choice  & Adam  & Adam & SGD\\
    \end{tabular}
    \caption{Hyperparameter Space. The hyperparmaters from the optimized phase and fluence neural networks are shown in their respective columns, along with the baseline architecture. }
    \label{tab:hp_table}
\end{table}
The neural network models were trained on a dataset of 1,830,000 generated samples, initially with $70\%$ of the data in the  training dataset  and $30\%$ in the validation data set. The networks were trained by gradient descent (backpropagation) using the training set. The validation dataset is then used to assess their performance and make sure the networks generalize properly to previously unseen data and do not overfit the training data. 

The input features to the neural network are the interpolated initial and SPM spectra, with a total of 140 features. The individual features of the input tend to be right-skewed, with a majority of events taking smaller scalar values and a small minority occurring in higher regions. In order to correct this we first take the log of the input features and then normalize them, by subtracting the mean and dividing by the standard deviation. The target variables are also normalized in the same fashion. Transforming the data through this process ensures all features are on the same scale.

Separate networks were trained to reconstruct the initial phase and the fluence of the pulse. The phase neural networks were trained with the targets being the initial phase of the pulse interpolated to the same 40 length frequency grid as the initial spectrum. The fluence networks were trained with the only target being the fluence of the initial pulse. All networks were implemented in Keras with a Tensorflow backend and trained on NVIDIA TITAN X GPUs.

All training samples were augmented with small amounts of Gaussian noise, $\mathcal{N}(0, 0.05)$, to mimic the imprecise fluctuations of experimental observations due to sources like laser fluctuations and thermal noise in silicon based detectors. This augmentation, added during training batches, also serves to prevent overfitting to the training set. Other models of experimental noise could be included by applying the noise model to the data, either during data generation or training. Training occurred over a maximum of 400 epochs.  The performance of each network is characterized by calculating the mean square error loss of the predicted values compared to the target values. If the validation loss did not improve after fifteen epochs, training was terminated. 

\begin{figure}[htbp]
\centering
\begin{subfigure}{\textwidth}
  \centering
  \includegraphics[width=0.75\linewidth]{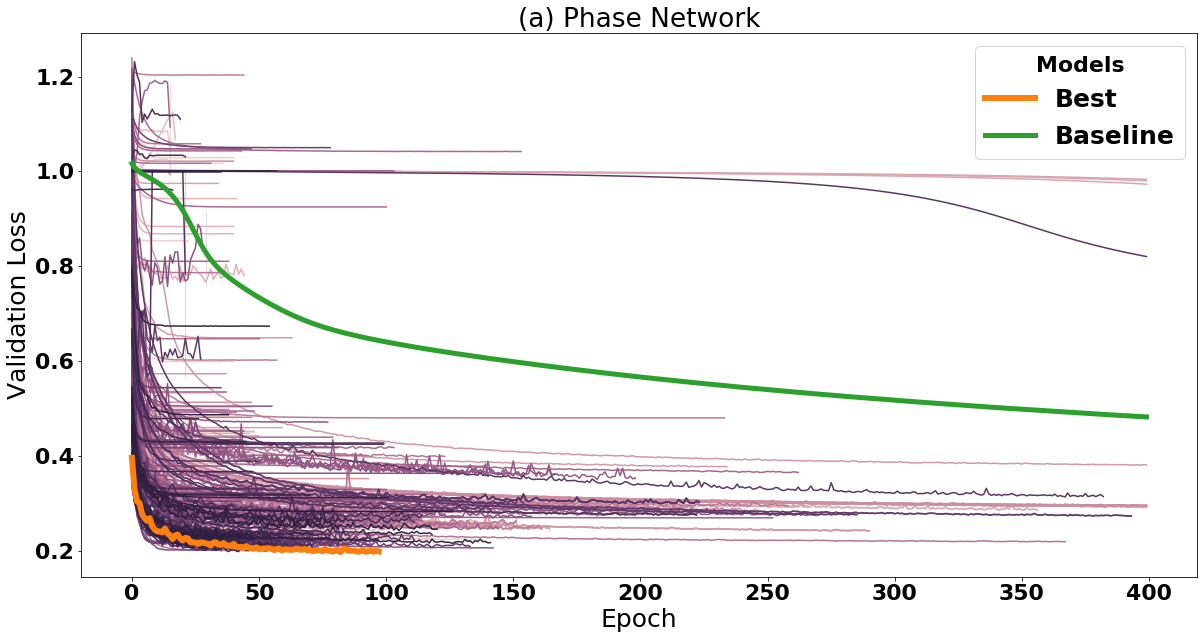}
  \label{fig:results}
\end{subfigure}\\

\begin{subfigure}{\textwidth}
  \centering
  \includegraphics[width=0.75\linewidth]{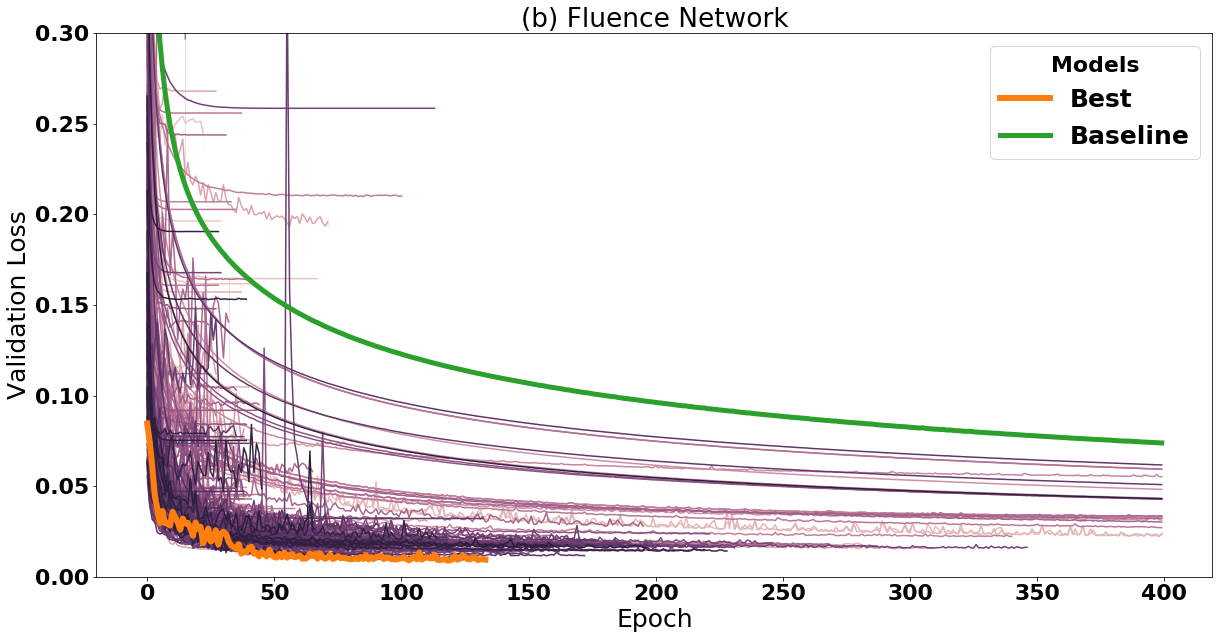}
  \label{fig:leaky_relu}
\end{subfigure}%

\caption{Validation loss of SHERPA trials, measured by the mean squared error, over time. Each line depicts the validation loss of a different SHERPA trial during the course of training. a) Trials from phase networks with varying hyperparameters. b) Trials from fluence networks with varying hyperparameters. Note: not all 500 trials are shown in each figure. Some trials with higher validation losses are left out for figure clarity. This discards 50 and 145 networks for a and b, respectively.}
\label{fig:sherpa_loss_plots}
\end{figure}

Building and training neural networks requires one to set many values, called hyperparameters, \textit{a priori}. Hyperparameters include the number of layers, the number of nodes per layer, the kinds of activation functions, the learning rates, and the dropout rates. Dropout is a randomization procedure used during training  that turns off different connections in the neural network, forcing the network to learn a more general solution which also helps avoid overfitting \cite{srivastava2014dropout,baldidropout14}.
In the experiments, the hyperparameters were optimized using SHERPA \cite{hertel2020}, a Python software library which is compatible with Keras and other modern deep learning libraries, and has been used to effectively optimize neural networks in various scientific applications (e.g. \cite{beucler2019enforcing, ott2020fortran}).

Leveraging SHERPA, a large suite of 500 models were explored using a Bayesian optimization algorithm. The Bayesian search has the advantage of learning a distribution over the hyperparameters of the network architecture, in relation to the task to be optimized. By employing this procedure we are able to evaluate a large space of possible models and test many configurations. To demonstrate the efficacy of the hyperparameter search, we compare the resulting model against an initially proposed baseline model. The baseline architecture is shown in Table \ref{tab:hp_table}. The optimized phase and fluence networks contained roughly 608 thousand and 970 thousand parameters, respectively.
\begin{figure}[ht]
\centering
\begin{subfigure}{\textwidth}
  \centering
  \includegraphics[width=0.75\linewidth]{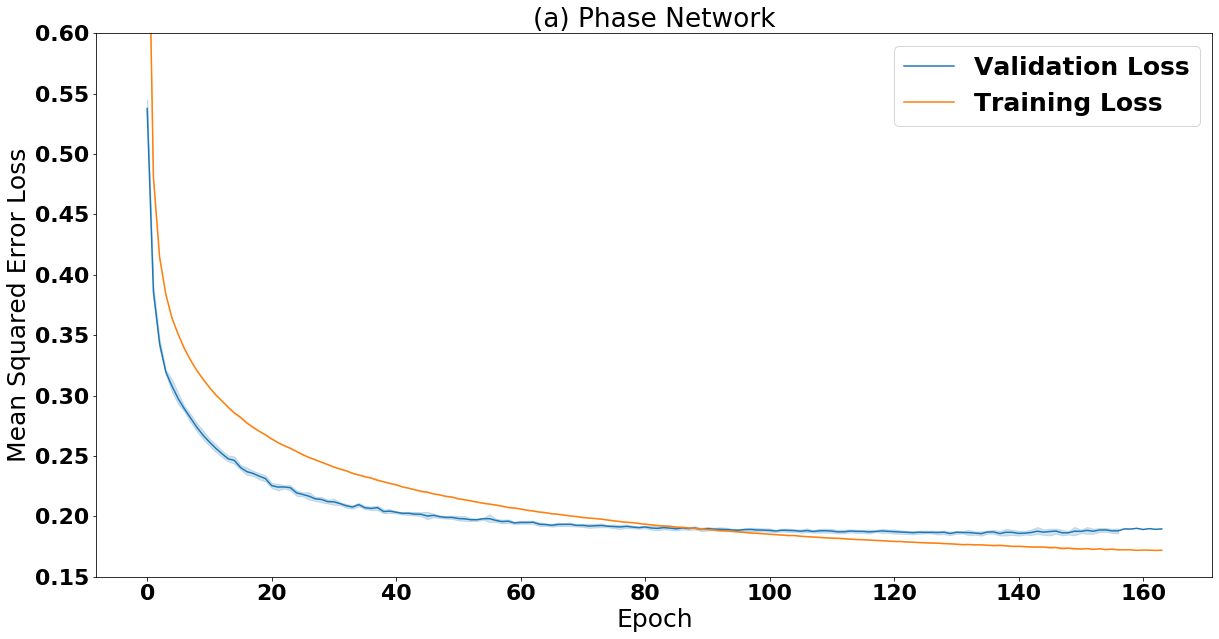}
  \label{fig:phase_crossval}
\end{subfigure}\\

\begin{subfigure}{\textwidth}
  \centering
  \includegraphics[width=0.75\linewidth]{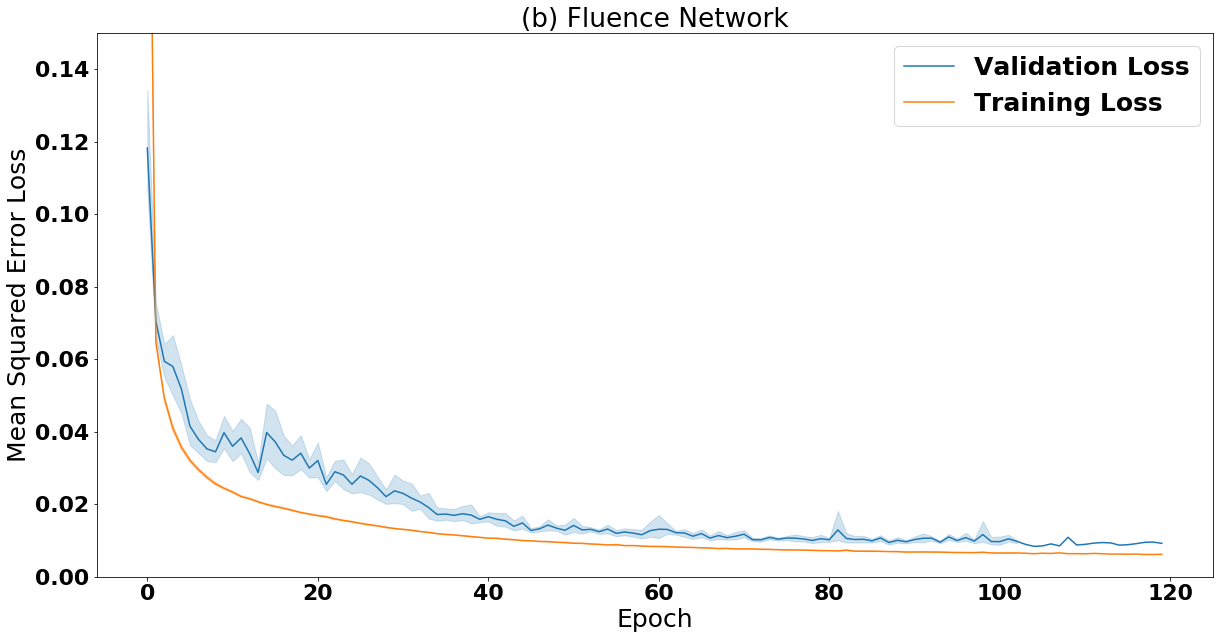}
  \label{fig:fluence_crossval}
\end{subfigure}%

\caption{Training and validation loss, measured by the mean squared error, over time. Loss curves show the average (solid line) and one standard deviation (shaded region) for the 10 folds of cross validation. a) Phase network b) Fluence network}
\label{fig:crossval}
\end{figure}
In total, 500 network architectures were explored with differing hyperparameters for both the phase and fluence neural networks. The final architectures from the hyper-parameter search are shown in Table \ref{tab:hp_table}. The table displays the hyperparameters of the best performing phase and fluence network, along with the hyperparameters of the baseline network. The distribution of the validation mean squared errors (MSE) for the phase and fluence networks are shown in Figure \ref{fig:sherpa_loss_plots}a and \ref{fig:sherpa_loss_plots}b respectively. These figures highlight the performance of the best optimized model compared to the initially proposed baseline network.

Following the hyperparameter search, the best performing phase and fluence networks were evaluated using 10-fold cross validation. During 10-fold cross validation the data is randomly partitioned into 10 distinct folds. Each network is then trained on 9 of the folds and tested on the remaining one, and the process is repeated 10 times. The mean and the standard deviation of the performance (error bars) can then be computed over the 10 experiments.  The results from 10-fold cross validation are presented in Figure \ref{fig:crossval}a and \ref{fig:crossval}b. These figures demonstrate consistent performance across all 10 folds. We confirm that neither the phase network nor the fluence network overfits the training data by comparing the performance on the training and validation set across all 10 folds. The average difference between the training and validation loss is less than 0.01 and 0.002 for the phase and fluence networks respectively.

\subsubsection*{Experimental Neural Networks}
 A neural network was then trained on 432 thousand pulses total using a 80/20 split for the training and validation data sets. Due the stability of the initial laser spectrum, the features used in training only needed to be based on the SPM spectrum. Both the broadened spectrum and the initial phase were interpolated to the range from 330 THz to 418 THz binned 100 linearly spaced bins. The network consisted of 8 layers with a width of 200 and was trained using a Gaussian noise of 0.1, learning rate off 0.001, drop out rate of 0.1 over 200 epochs. The phases reconstruction from the neural network were then compared to the phase reconstructed from a second harmonic generation FROG measurement of the initial beam as shown in Fig \ref{fig:experimental}. 

\section*{Acknowledgements}
This work was supported by STROBE: A National Science Foundation Science \& Technology Center under Grant No. DMR-1548924; This material is based upon work supported by the National Science Foundation under the CAREER program Grant No. PHY-1753165, and under grant number DGE-1633631.

\section*{Data availability}
The data related to this paper is available from the corresponding author upon request.

\section*{Contributions}
M. S., J. O., D. F., C. M., P. B., and F. D. contributed to the initial concept of the project. M. S. generated the simulated data required for training of the networks. J. O. and P.B designed, trained, optimized, and validated the neural networks for the broadband dataset. M. S. trained the neural network for the experimental dataset. M. S., C. G., and N. B. collected the experimental data. F. D. and P. B. supervised the project. M. S. and J. O. prepared the manuscript. All authors edited the paper.

\bibliographystyle{unsrt}
\bibliography{main}

\end{document}